\documentclass{elsart}
\usepackage{amsmath}
\usepackage{latexsym}
\usepackage{float}
\usepackage{amssymb}
\usepackage{graphicx}
\usepackage{times}
\usepackage{bm}
\usepackage{setspace}

\begin{document}

\begin{frontmatter}

\title{Glassy behavior of molecular crystals: A comparison between
results from MD-simulation and mode coupling theory}

\author{M.~Ricker$^{a}$, F.~Affouard$^{b}$, R.~Schilling$^{a}$ and M.~Descamps$^{b}$}

\address[$^{a}$]{Institut f\"ur Physik, Johannes Gutenberg-Universit\"at
  Mainz, Staudinger Weg 7, D-55099 Mainz, Germany}
\address[$^{b}$]{
Laboratoire de Dynamique et Structure des Mat\'eriaux Mol\'eculaires,
UMR CNRS 8024,
Universit\'e Lille I,
59655 Villeneuve d'Ascq Cedex France}

\maketitle

\begin{abstract}
We have investigated the glassy behavior of a molecular crystal
built up with chloroadamantane molecules. For a simple model of
this molecule and a rigid fcc lattice a MD simulation was
performed from which we obtained the dynamical orientational
correlators $S_{\lambda \lambda '}({\bf{q}},t)$ and the ``self''
correlators $S_{\lambda \lambda '}^{(s)}(t)$, with $\lambda =
(\ell ,m)$, $\lambda' = (\ell' ,m')$. Our investigations are for
the diagonal correlators $\lambda = \lambda'$. Since the lattice
constant decreases with decreasing temperature which leads to an
increase of the steric hindrance of the molecules, we find a
strong slowing down of the relaxation. It has a high sensitivity
on $\lambda$, $\lambda '$. For most $(\ell,m)$, there is a
two-step relaxation process, but practically not for $(\ell,m) =
(2,1)$, $(3,2)$, $(4,1)$ and $(4,3)$. Our results are consistent
with the $\alpha$-relaxation scaling laws predicted by mode
coupling theory from which we deduce the glass transition
temperature $T_c^{MD} \cong 217K$.  From a first principle
solution of the mode coupling equations we find $T_c^{MCT} \cong
267K$. Furthermore mode coupling theory reproduces the absence of
a two-step relaxation process for $(\ell,m)=(2,1)$, $(3,2)$,
$(4,1)$ and $(4,3)$, but underestimates the critical nonergodicity
parameters by about $50$ per cent for all other $(\ell,m)$. It is
suggested that this underestimation originates from the
anisotropic crystal field which is not accounted for by mode
coupling theory. Our results also imply that phonons have no
essential influence on the long time relaxation.
\end{abstract}


\end{frontmatter}

\section{Introduction}
The mode coupling theory (MCT) \cite{1} is the most successful
\textit{microscopic} approach to describe glassy dynamics of
supercooled liquids on a qualitative, and partly even on a
quantitative level \cite{2,3,4}. Particularly,
\textit{first-principle} comparisons between the solutions of MCT
equations and results from experiments and simulations for various
systems like the binary Lennard-Jones liquid \cite{5}, liquids of
diatomic molecules \cite{6,7}, water \cite{8,9}, silica melt
\cite{10} and a model for orthoterphenyl \cite{11} have confirmed
the quality of MCT. These comparisons were restricted to the glass
order parameters, the nonergodicity parameters. Going beyond that,
even time dependent quantities, like the intermediate scattering
function, have been compared with each other for the binary
Lennard-Jones liquid \cite{12}, binary mixtures of hard spheres
\cite{13,14} and the polydisperse quasi-hard-sphere system \cite{15},
again demonstrating consistency with MCT.

Glassy dynamics of systems with self-generated disorder is not
restricted to liquids. Molecular crystals in their plastic phase
exhibit glassy behavior as well. This has been found
experimentally three decades ago \cite{16}. Recently two of the
present authors have extended MCT to molecular crystals \cite{17}.
Using the static structure factors from Percus-Yevick theory
\cite{18} the orientational glass transition of uniaxial hard
ellipsoids on a simple cubic lattice has been investigated
\cite{17}. There it has been found that this transition is not
driven by an orientational cage effect, analogous to supercooled
liquids, but by the growth of orientational order. This growth
manifests itself in an increase of the \textit{static}
orientational correlators at the Brillouin center or edge, which
in turn leads to an increase of the memory kernel. However, before
the corresponding orientational correlation length diverges at the
corresponding equilibrium phase transition line, the nonlinear
feedback mechanism of MCT results in an orientational glass
transition. The same mechanism has been identified for a liquid of
uniaxial hard ellipsoids \cite{19}. Accordingly, the MCT glass
transition line is located within the orientationally disordered
ergodic phase \cite{17,19} and not in the supercooled regime.

One may ask whether there also exists a cage-effect-driven glass
transition for molecular crystals. Good candidates are plastic
crystals which undergo a first order equilibrium phase transition
to an orientationally ordered phase. Several such systems exist
such as cyanoadamantane, ethanol, cyclooctanol, difluorotetrachloro\-ethane
or $\mathrm{C_{60}}$~\cite{Affouard_jncs02}. Another one is chloroadamantane,
which exhibits an equilibrium phase transition at $T^{exp}_{eq} \cong 244K$
\cite{Foulon_actacryst89}. A simple
model for chloroadamantane (see next section) has been studied by
MD simulations \cite{20,21}. A MCT analysis for the critical
amplitudes and the $\alpha$-relaxation time have demonstrated
consistency with MCT predictions \cite{21}. These predictions can
really be extended from liquids to plastic crystals because the MCT
equations for the latter have the same mathematical structure than
for multi-component simple liquids \cite{17}.

It is the main goal of the present paper to perform a first
principle comparison between the results for chloroadamantane from
a MD-Simulation and MCT. The outline of our contribution is as
follows. In the second section we will describe the
model, introduce the orientational correlation functions,
give some details of the simulational procedure and
shortly describe the relevant MCT equations. Results are presented
and discussed in the third section and the final section contains a
summary and some conclusions.

\section{Model and technical details}

Chloroadamantane $C_{10}H_{15}Cl$ is a rather huge molecule which belongs to
the substituted adamantane family.  It shows a plastic phase structure
isomorphous to cyano\-adamantane, but the chloroadamantane molecule possesses a
smaller substitute and a faster dynamics well adapted for MD simulation
investigations.  Chloroadamantane undergoes at $T \simeq 244K$ a first order
transition from an ordered monoclinic structure to a rotator phase with
face-centered-cubic (fcc) symmetry \cite{Foulon_actacryst89}.  The
plastic-liquid transition occurs at $T_{m} \simeq 442K$ \cite{Foulon_actacryst89}.

\begin{center}
\begin{table}[h!]
\caption{Parameters for the two-site chloroadamantane model.}

\label{table1}
\begin{tabular}{|l|c|c|c|c|}
\hline
Site-Site & p & q & $\epsilon$ (kJ/mol) & $\sigma$ (\AA)  \\ \hline
Cl -  Cl  & 12  &  6 & 1.441 &  3.350 \\ \hline
Cl -  Adm  & 14  &  8 & 3.087 & 4.786   \\ \hline
Adm  -  Adm  & 16  & 11 & 12.47 & 6.200   \\ \hline
\end{tabular}
\end{table}
\end{center}
The simulated system
is composed of rigid linear molecules
with two sites:
one chlorine atom (noted Cl) and
one super atom (noted Adm) that models
the adamantane part $\mathrm{C_{10}H_{15}}$.
The moment of inertia is 302.733 $\mathrm{amu.\AA^{2}}$.
Molecular dynamics calculations were performed on a
system of $N = 256$ molecules
($4 \times 4 \times 4$ fcc crystalline cells)
interacting
through a Lennard-Jones short range site-site potential
of the form
$$
v(r) = 4 \epsilon
\left( (\sigma/r)^{p}
- (\sigma/r)^{q}
\right)
$$
where $r$ is the distance between two different sites. The parameters
$\epsilon$, $\sigma$, $p$ and $q$ are specified in table~\ref{table1}.

The chloroadamantane molecule possesses a relatively large dipolar moment
$\vec{\mu}$ (2.39 Debyes) which is parallel to the molecular axis.  The
electrostatic interactions were handled by the Ewald method with two partial
charges ($q= \pm 0.151e$) localized on both sites.  Newton's equations of
motion were solved with a time step of $\Delta t = 5$ fs. We worked in the NPT
(constant number of molecules, temperature and pressure)  and NVT  (constant
number of molecules, temperature and volume) statistical ensembles with
periodic boundaries conditions.  The sample was first equilibrated in the NPT
ensemble. Then, MD runs at constant volume using the average volume determined
from the NPT simulations were performed.

Since our molecule-model has one rotational symmetry
axis we can use the angles $\Omega_n(t)=(\Theta_n(t),\phi_n(t))$ to
characterize the orientation of the n-th molecule at site $n$ and
time $t$. The microscopic local orientational density is given by:
\begin{equation}\label{eq1}
\rho_n(\Omega,t)=\delta(\Omega|\Omega_n(t))
\end{equation}
with $\delta(\Omega|\Omega  ')=(\sin \Theta)^{-1} \delta (\Theta -
\Theta ')\delta(\phi - \phi ')$. Expansion with respect to
spherical harmonics $Y_\lambda (\theta,\phi),\; \lambda=(\ell,m)$
and performing a lattice Fourier transform leads to the
tensorial orientational density modes:
\begin{equation}\label{eq2}
\rho_\lambda({\bf{q}},t)=i^l\,\sum \limits _{n=1}^N Y_\lambda
(\Omega_n(t)) e ^{i{\bf{qR}}_n}
\end{equation}
where $N$ is the number of lattice sites and $\bf{R}_n$ the lattice
vector of the $n$th lattice site. The reader should note that we
assume a \textit{rigid} lattice and that ${\bf{q}}$ is restricted
to the 1. Brillouin zone. Introducing the fluctuations $\delta
\rho _\lambda({\bf{q}},t)=\rho_\lambda ({\bf{q}},t)-\langle
\rho_\lambda ({\bf{q}},t)\rangle$ we can define the time
dependent, tensorial orientational correlators:
\begin{equation}\label{eq3}
S_{\lambda \lambda '}({\bf{q}},t)= \frac {4 \pi}{N} \langle \delta
\rho _\lambda ^*({\bf{q}},t)\delta \rho_{\lambda '}
({\bf{q}},0)\rangle
\end{equation}
where $\langle \ldots \rangle$ denotes the canonical average over
the initial conditions. Note that the absence of a head-tail
symmetry leads to nontrivial correlators for all $\ell \geq 1$,
while the absence of phonons leads to $S_{\lambda \lambda
'}({\bf{q}},t) = 0$ for $\lambda = (0,0)$ and/or $\lambda' =
(0,0)$ within the 1. Brillouin zone.

\begin{figure}[t]
\centerline{\includegraphics[width=6.5cm,angle=270,clip=true]{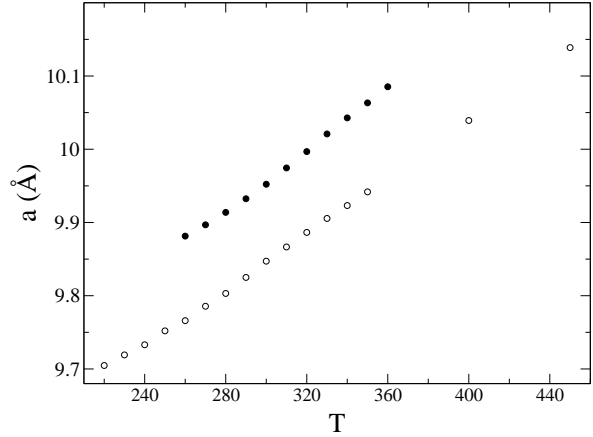}}
\caption{Temperature dependence of the lattice
constant obtained from experiments (full circles) and
MD-simulation (open circles) \label{fig1}}
\end{figure}
These correlators have been investigated by a MD-simulation. In
contrast to earlier work \cite{20,21} the present simulation has
been done for each temperature $T$ for a rigid lattice with
lattice constant $a(T)$ determined from the average size of
the system obtained from NPT simulations. It is
represented in Figure~\ref{fig1}, which also contains the experimental
result \cite{Foulon_actacryst89}. It is the shrinking of
the lattice constant (cf. Fig.~\ref{fig1}) which leads to an
increase of the steric hindrance with decreasing temperature and
in turn to a slowing down of the orientational dynamics.

On the other hand, MCT provides an equation of motion for
$S_{\lambda \lambda '}({\bf{q}},t)$ which requires the static
correlators $S_{\lambda \lambda '}({\bf{q}},0)$ as an input.
Because of the crystal's anisotropy, these correlators also depend
on the direction of ${\bf{q}}$. Therefore the numerical solution
of the t-dependent MCT equations is rather involved such that we
will restrict ourselves to the determination of the MCT glass
transition temperature $T_c$ and the normalized nonergodicity
parameters
\begin{eqnarray}\label{eq4}
f_{\lambda \lambda '}({\bf{q}}) = \lim _{t \rightarrow \infty}
\phi_{\lambda \lambda '} ({\bf{q}},t), \\ \nonumber \phi_{\lambda
\lambda '} ({\bf{q}},t)= S_{\lambda \lambda '}({\bf{q}},t)/[S _{\lambda
\lambda}({\bf{q}},0)S_{\lambda ' \lambda '}({\bf{q}},0)]^{1/2}
\end{eqnarray}
taken at $T=T_c$. They are solutions of an infinite set of
nonlinear coupled algebraic equations \cite{17}:
\begin{equation}\label{eq5}
f_{\lambda \lambda '}({\bf{q}}) = {\mathcal{T}}_{\lambda \lambda
'} ({\bf{q}},\{f_{\lambda \lambda '}({\bf{q}})\})\quad .
\end{equation}
The mode coupling polynomial ${\mathcal{T}}_{\lambda \lambda '}$
is related to the memory kernel and depends on temperature through
the static correlators. The static correlators have been taken
from the MD-simulation. The numerical solution of Eq.~(\ref{eq5})
requires a truncation at $\ell_{max}$, for which we have chosen
$\ell_{max} = 4$.

\section{Results}
\begin{figure}[t]
\centerline{\includegraphics[width=6.5cm,angle=270,clip=true]{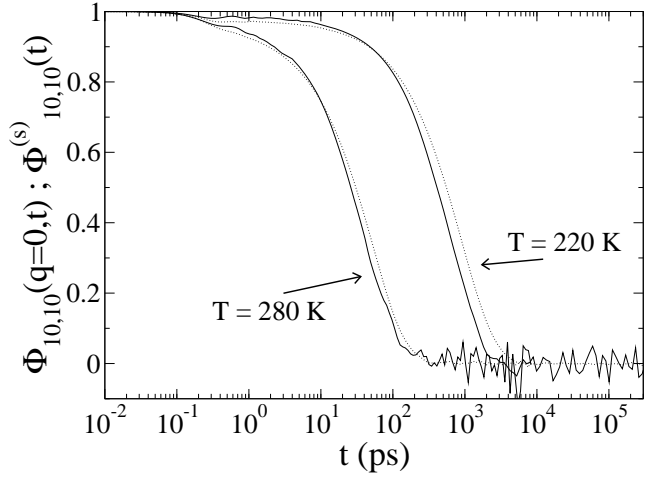}}
\caption{Time-dependence of the collective
correlator $({\bf{q}} = {\bf{0}})$  (solid line) and ``self''
correlator (dotted line) for $\lambda = \lambda ' =(1,0)$ and $T =
220 K, 280 K$ \label{fig2}}
\end{figure}
\begin{figure}[t]
\centerline{\includegraphics[width=7.0cm,angle=0,clip=true]{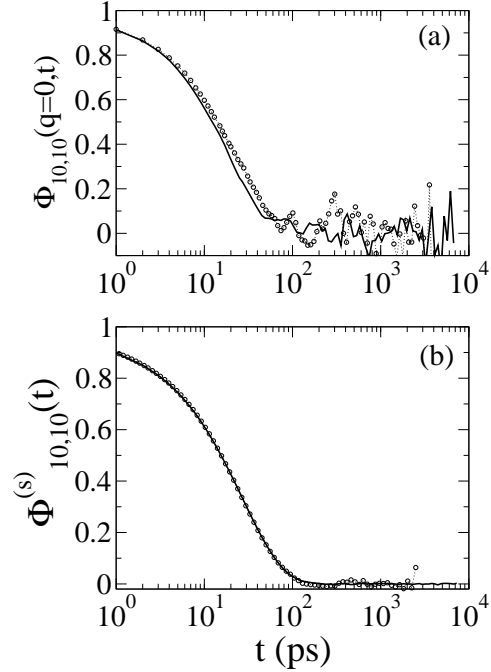}}
\caption{Time-dependence of (a) the collective
correlator $({\bf{q}} = {\bf{0}})$ and (b) ``self'' correlator for
$\lambda = \lambda'=(1,0)$ at $T=300 K$ for $N=256$ particles
(open circles) and $N=2048$ particles (solid line), respectively.
\label{fig3}}
\end{figure}

As mentioned above the collective correlators (Eq.~(\ref{eq3}))
are nontrivial for $\ell \geq 1, \ell ' \geq 1$. Besides these,
one can also determine the ``self'' correlators:
\begin{eqnarray}\label{eq6}
S^{(s)}_{\lambda \lambda '}(t) &=& \frac {4 \pi}{N} \sum \limits
^N _{n=1} \langle \delta \rho _{n, \lambda}^*(t) \delta \rho
_{n,\lambda '}(0)\rangle \; \\ \nonumber &=& \frac{1}{N}\sum \limits
_{{\bf{q}} \in 1.B.Z.}S_{\lambda \lambda '}({\bf{q}},t).
\end{eqnarray}
$\delta \rho_{n,\lambda}(t)$ is the expansion coefficient of $\delta
\rho_n(\Omega,t)$ with respect to $Y_\lambda(\Omega)$. Figure~\ref{fig2}
shows the normalized collective correlator at ${\bf{q}}={\bf{0}}$ and
the ``self'' correlator for $\lambda = \lambda '= (1,0)$ and $T= 220
K$ and $280K$. First of all we observe a slowing down of the
$\alpha$-relaxation by a factor of about twenty by changing the
temperature from $280K$ to $220K$, i.e. by about one fifth. Second,
no significant difference occurs between the collective and the
``self'' correlator. The former possesses some oscillations on the
microscopic time scale below 10 ps, in contrast to the self
correlator. At the lowest temperature $220K$ both reveal a two-step
relaxation process, however, with a plateau height very close to
one. Since the system size of $256$ sites is not very huge, we have
also performed a simulation at $T = 300K$ with $2048$ sites. The
result for $\lambda = \lambda '=(1,0)$ is presented in Figure~\ref{fig3}a
(collective) and Figure~\ref{fig3}b (``self'') and shows no significant
finite size effects. Whether this holds also at the lowest
temperature $T = 220K$ has not been possible to study because
of very long equilibration times.

\begin{figure}[t]
\centerline{\includegraphics[width=7.0cm,clip=true]{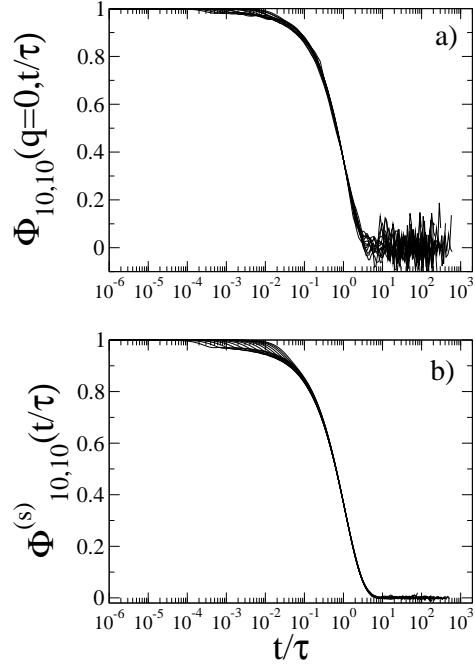}}
\caption{(a) Collective correlator $({\bf{q}} =
{\bf{0}})$ and (b) ``self'' correlator for $\lambda = \lambda ' =
(1,0)$ versus the rescaled time $t/\tau$. $\tau$ has been
determined as the time at which the correlators have decayed to
$1/e$. Correlators are shown for $T= 220K$ to $350K$ in $10K$ steps.
\label{fig4}}
\end{figure}
\begin{figure}[tbh]
\centerline{\includegraphics[width=6.5cm,angle=270,clip=true]{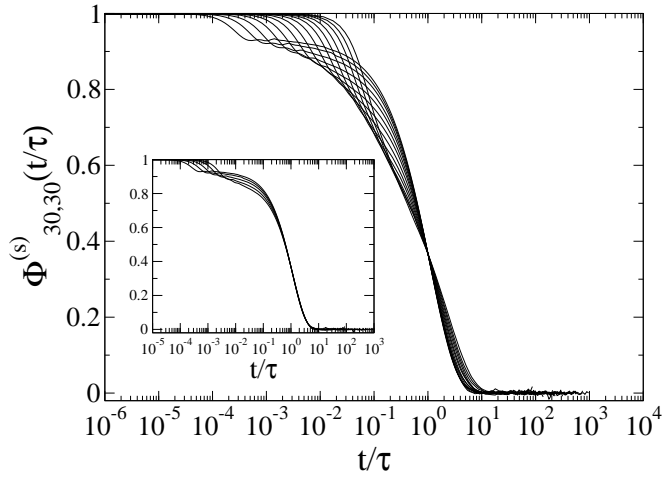}}
\caption{Same as Fig.\ref{fig4}b, but for $\lambda = \lambda
' =(3,0)$. The inset shows the lowest five temperatures, only.
\label{fig5}}
\end{figure}
\begin{figure}[t]
\centerline{\includegraphics[width=7.0cm,angle=0,clip=true]{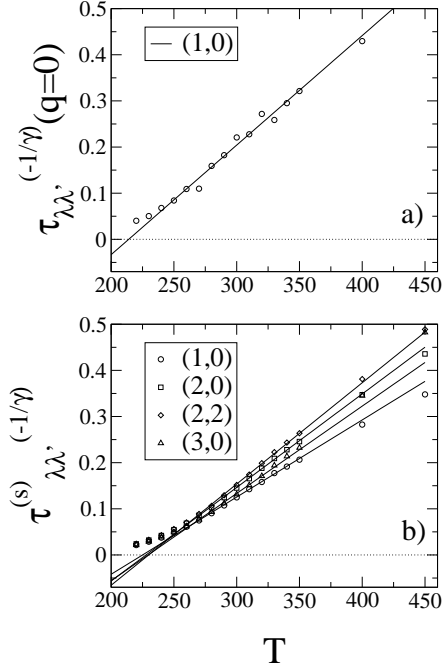}}
\caption{$\tau^{-1/\gamma}$ versus $T$ for (a)
the  collective correlator for $\lambda = \lambda ' = (1,0)$
at ${\bf{q}} = {\bf{0}}$ and (b) the ``self'' correlator for
$\lambda = \lambda ' = (1,0),(2,0),(2,2)$ and $(3,0)$. \label{fig6}}
\end{figure}
One of the predictions of MCT is the validity of the $t-T$
superposition principle, i.e. scaling of t by the
$\alpha$-relaxation time $\tau_{\lambda \lambda'}({\bf{q}},T)$
should yield a data collapse for all $T$ and $t$ large enough.
Taking $\tau_{\lambda \lambda '}({\bf{q}},T)$ as the time at which
$\phi_{\lambda \lambda '}({\bf{q}},t)$ has decayed to $1/e$ the
rescaled data are given in Figure~\ref{fig4}a and Figure~\ref{fig4}b
for the collective $({\bf{q}}={\bf{0}})$ and ``self'' correlator,
respectively, for $\lambda = \lambda '=(1,0)$. A satisfactory
collapse is found over a rather huge temperature range. As can be
seen from Figure~\ref{fig5} this is not true for $\phi_{30,30}^{(s)}(t)$.
Restriction to $220K \leq T \leq 260K$ leads to a satisfactory result
for $t/\tau >1$, which, however, excludes the von Schweidler regime
(see inset of Fig.~\ref{fig5}). One of the highly nontrivial
predictions of MCT is the validity of the power law dependence
of $\tau$ on $(T-T_c)$:
\begin{equation}\label{eq7}
\tau (T)\sim (T-T_c)^{-\gamma} \quad , \quad T \geq T_c
\end{equation}
Figure~\ref{fig6}b shows $\tau^{-1/\gamma}$ versus $T$ for the ``self''
correlators for different $\lambda = \lambda '$ and Figure~\ref{fig6}a
the corresponding result for the collective one at ${\bf{q}}=
{\bf{0}}$ and $\lambda = \lambda '=(1,0)$. There is a linear
behavior for $250K <T<350K$. Its extrapolation to zero yields
\begin{equation*}
T_c^{MD} \cong 217 \pm 5 \,\, K
\end{equation*}
For the exponent we find $\gamma_{\lambda \lambda '} \cong 2.0 \pm 0.2 $
depending on $\lambda =\lambda'$.

\begin{figure}[!h]
\centerline{\includegraphics[width=7.0cm,angle=0,clip=true]{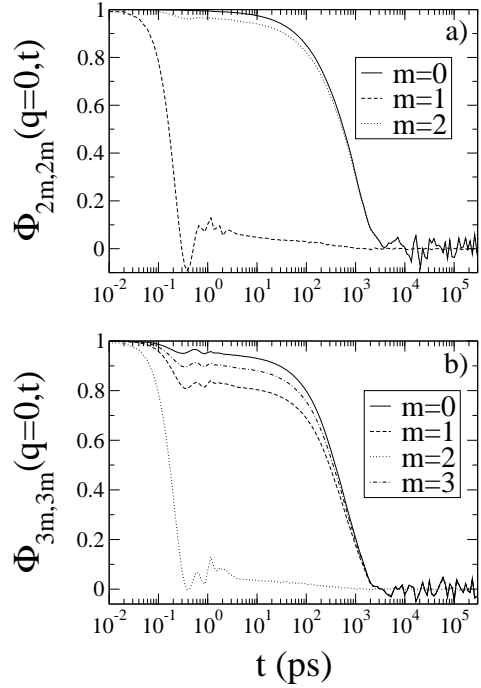}}
\caption{Time- and m-dependence of the collective
correlator at ${\bf{q}} = {\bf{0}}$ and $T=$ 220 K
for (a) $\ell = \ell '=2$ and (b) $\ell = \ell'=3$
\label{fig7}}
\end{figure}
An interesting behavior has been found for the collective
correlators with $\lambda = \lambda '$ and $\ell = 2,3$. The
corresponding results at the \textit{Brillouin center} are
presented in Figure~\ref{fig7}a and~\ref{fig7}b for the lowest
temperature $T=220K$. Both correlators exhibit a high sensitivity
on the $m$-index. For $\ell =2$ and $\ell =3$ there is a two-step
relaxation for $m=0,2$ and $m=0,1,3$, respectively, with rather
high plateau values, particularly for $\ell =2$. But for $\ell =2,
m=1$ and $\ell =3,m=2$ there is a rather unusual t-dependence
showing no typical two-step relaxation with a convex ``short'' and
a concave ``long'' time part joining at an inflection point (the
$\beta$-relaxation regime of MCT). The ``long'' time relaxation
for $t>4 \,ps$ has a convex curvature on the logarithmic time
scale starting at a height of about 0.05 for both cases. If this
height is interpreted as a plateau height, it has an unusual small
value. Surprisingly, it is observed that the correlators without
two step relaxation decay faster as the temperature is lowered.

\begin{figure}[tbh]
\centerline{\includegraphics[width=8.0cm,angle=0,clip=true]{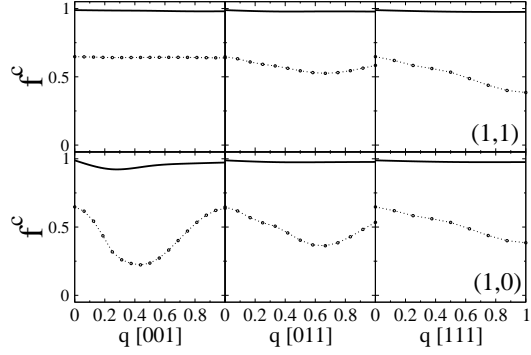}}
\caption{${\bf{q}}$-dependence of the critical
nonergodicity parameters along the [001], [011] and [111]
directions for $\lambda = \lambda '=(1,0),(1,1)$. The result from
the MD simulation is given by the solid line and from MCT by the
dots (the line connecting the dots are a guide for the eye)
\label{fig8}}
\end{figure}
For  $(\ell,m) = (4,0)$, $(4,2)$ and $(4,4)$ we also have a two
step relaxation, while for $(\ell,m) = (4,1)$ and $(4,3)$
it is missing, as for $(2,1)$, $(3,2)$. The results for $\ell = 4$
yield no new physical insight and are not shown in this work.

\begin{figure}[tbh]
\centerline{\includegraphics[width=8.0cm,angle=0,clip=true]{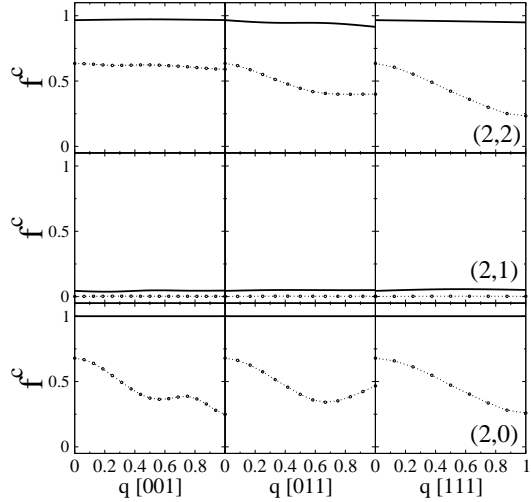}}
\caption{Same as Fig.~\ref{fig8}, but for $\lambda = \lambda
'= (2,0),(2,1),(2,2)$ \label{fig9}}
\end{figure}
In a final step we have investigated how far these plateau values
(which are roughly equal to the critical nonergodicity parameters)
are reproduced by MCT for plastic crystals. First, we have
fitted the MD-results by the von Schweidler law
in order to determine the critical nonergodicity parameters
$f_{\lambda \lambda '}({\bf{q}})$ from the correlators at the
lowest simulated temperature $T=220 K$, which is close
to $T_c^{MD}$. On the other hand we have used the static
correlators from the simulation to solve Eq.~(\ref{eq5}) by
iterations. This allows to locate the MCT-temperature $T_c^{MCT}$,
at which the trivial solution $f_{\lambda \lambda '}({\bf{q}})
\equiv 0$ changes discontinuously to $f_{\lambda \lambda'}
({\bf{q}}) \neq 0$.

\begin{figure}[t]
\centerline{\includegraphics[width=8.0cm,angle=0,clip=true]{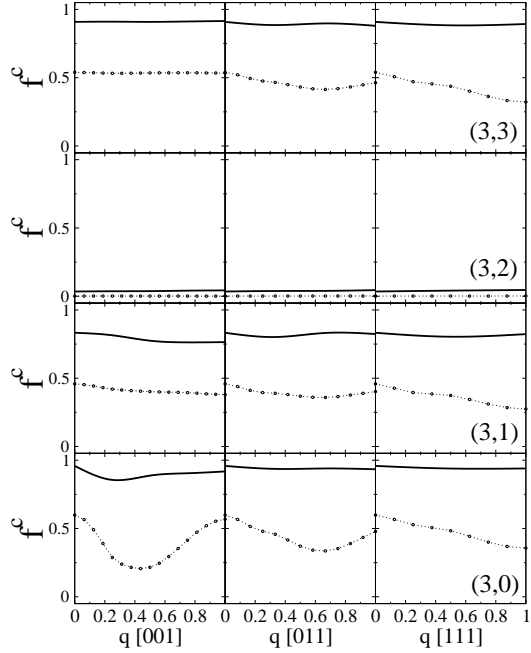}}
\caption{Same as Fig.~\ref{fig8}, but for $\lambda = \lambda
' =(3,0),(3,1),(3,2),(3,3)$ \label{fig10}}
\end{figure}
As a result we have found
\begin{equation*}
T_c^{MCT} \cong 267K
\end{equation*}
The corresponding critical nonergodicity parameters from MCT are
given together with those from the simulation in
Figures~\ref{fig8}-\ref{fig10}. The q-dependence is shown along
three highly symmetric reciprocal directions. Let us discuss first
$\ell = \ell ' =1$ (Fig.~\ref{fig8}). The numerical result is very
close to one for all q. For $m=m'=0$, there is a shallow minimum
along [001] direction. This minimum also appears in the
corresponding MCT-result. However, $f_{10,10}^c({\bf{q}})$ from
MCT is \textit{not} close to one and deviates from the MD-result
by about $50$ per cent. Furthermore, the former exhibits a much
stronger q-dependence, whereas the MD-results are practically
q-independent with exception for $m=m'=0$ along the $[001]$
direction. Similar conclusions can be drawn for $\ell = \ell ' =2$
(Fig.~\ref{fig9}) and $\ell = \ell ' =3$ (Fig.~\ref{fig10}). There
are two features which are reproduced by MCT. \textit{First}, we
observe that the MCT-result for $f^c_{21,21}({\bf{0}})$
(Fig.~\ref{fig9}) and $f^c_{32,32}({\bf{0}})$ (Fig.~\ref{fig10})
is practically zero, in semi-quantitative agreement with the
corresponding MD-result. This is even true for all ${\bf{q}}$.
\textit{Second}, the hierarchy $f^c_{21,21}({\bf{0}})
<f_{22,22}^c({\bf{0}}) <f^c_{20,20}({\bf{0}})$ and
$f_{32,32}^c({\bf{0}}) <
f_{31,31}^c({\bf{0}})<f_{33,33}^c({\bf{0}})<f_{30,30}^c({\bf{0}})$
is valid for both MCT and MD.

\section{Summary and conclusions}
The main motivation of the present paper has been the test of the
validity of the MCT predictions for the glassy behavior of
molecular crystals. MCT has been recently extended from liquids to
molecular crystals \cite{17}. Since the corresponding equations of
motion have the same structure as for multicomponent simple
liquids, the MCT predictions for the liquid systems also hold for
molecular crystals.

In order to check the validity of these predictions we have
performed a MD simulation for a simple model of chloroadamantane.
In contrast to earlier simulations \cite{20,21}, we have used a
\textit{rigid} fcc-lattice, however, with an appropriately chosen
temperature-dependent lattice constant $a(T)$. The decrease of
$a(T)$ with decreasing temperature enhances the steric hindrance,
which is responsible for the glassy dynamics. Figure~\ref{fig2}
clearly demonstrates the validity of this physical picture.

There are two kinds of tests. On the one hand one can check the
validity of the two scaling laws predicted by MCT \cite{2,3,4}.
Without attempting to calculate the corresponding exponents from
first principles this represents an important qualitative test. On
the other hand a first principle comparison can be made between
the MD- and MCT-results, as has been done in
Refs.~\cite{5,6,7,8,9,10,11,12,13,14,15}. Concerning the test of
the scaling laws we have restricted ourselves to the second
scaling law which holds for the $\alpha$-relaxation regime. In
this regime, MCT for liquids and molecular crystals predicts the
$t-T$ superposition principle which implies a data-collapse under
rescaling time by the $\alpha$-relaxation time $\tau$.
Figure~\ref{fig4} clearly demonstrates the validity of the second
scaling law over a large temperature range for e.g.~the collective
and ``self'' correlators with $\lambda = \lambda' = (1,0)$.
Increasing $\ell$ and $\ell '$ leads to a shrinking of that
temperature range and also of the interval of validity for
$t/\tau$ (see Fig.~\ref{fig5}). The T-dependence of $\tau$ is
shown in Figure~\ref{fig6}. As predicted by MCT a power law
dependence is found with a glass transition temperature $T_c^{MD}
\cong 217K$.
 Close to $T_c^{MD}$ deviations from the power law exist,
due to ergodicity restoring processes. The value $T_c^{MD}$ is
about $20$ per cent below $T_c^{MCT}$, consistent with what has
been found for liquids \cite{5,6,7,8,9,10,11}.

The MD-results have shown a very high sensitivity of the
relaxational behavior on $m$. For instance the collective
correlators for $\ell = \ell ' =2$ and $m =m'=0,2$ exhibit a
two-step relaxation process with a plateau very close to one, but
a very peculiar behavior for $m=m'=1$ (see Fig.\ref{fig7}a). For
the latter there is practically no plateau. The same holds for
$\ell = \ell ' =3$ and $m=m'=2$ (see Fig.~\ref{fig7}b). A first
principle comparison between the MD- and MCT-results (see
Figs.~\ref{fig8}-\ref{fig10}) reproduces this unusual
m-dependence. We stress that this unusually low plateau value is
\textit{not} related to a small corresponding static correlator.
However, the critical nonergodicity parameters for $\ell = \ell
=2$, $m=m'=0,2$ and for $\ell = \ell ' =3$, $m=m'=0,1,3$ from MCT
deviate by about $50$ per cent from the numerical ones. This
strong discrepancy can have several reasons. First, the truncation
at $\ell_{max}= 4$ may lead to a significant underestimation of
the critical nonergodicity parameters. Second, there may be quite
a different reason we want to explain now. The potential energy of
the rigid molecular crystal can be written as follows:
\begin{equation}\label{eq8}
V(\Omega_1,\ldots,\Omega_N)=V_0+ \sum\limits_n V_1(\Omega_n)+
\frac 1 2 \sum \limits _{n \neq m} V_2 (\Omega _n, \Omega _m)
\end{equation}
in case of two-body interactions. The isotropic part $V_0$ is not
essential. Besides the pair interactions given by $V_2$ there is a
one-particle term $V_1$, which is a kind of crystal field, due to
the crystal's anisotropy. $V_1(\Omega)$ will have a finite number
of local minima. If the pair interaction $V_2$ would be zero, the
molecules at finite temperatures would perform thermally activated
jumps between these local minima, as can be really seen in Figure
2 of Ref.~\cite{20}. Starting in one of the local minima it is
obvious that the long time limit of, e.g. the ``self''
correlators, will be positive. This happens for all initial
conditions with energy below the corresponding energy barrier, and
has already been investigated for a $\phi^4$-model \cite{22}. This
type of cage effect generated by an anisotropic one-particle
potential is \textit{not} accounted for by MCT which describes a
two-, three-, etc. particle-cage-effect in a self consistent way.
We believe that the large numerical values for the critical
nonergodicity parameters are mainly due to such a
one-particle-cage-effect.

Let us finally comment on the role of phonons, which are absent in
the present MD-simulation, and also in MCT for molecular crystals
\cite{17}. A comparison of, e.g. $T_c^{MD}$ and $\gamma$ from the
present simulation with earlier ones for the same model for
chloroadamantane but on a non-rigid lattice shows that phonons do
not have a noticeable influence on these quantities and on the
long time dynamics.

\bigskip

{\bf Acknowledgments}

Two of the authors (F. A. and M. D.)
wish to acknowledge the use of the facilities of the IDRIS (Orsay, France) and
the CRI (Villeneuve d'Ascq, France) where calculations were carried out.
This work was supported by the INTERREG III (FEDER)
program (Nord-Pas de Calais/Kent).

%
%
%
%
%
%
%
%
%
%
%
%
%

\end{document}